\documentclass{singlecol}

\usepackage{natbib,stfloats}
\usepackage{mathrsfs}
\usepackage{array}
\usepackage{graphicx}
\usepackage[english]{babel}
\usepackage{color}
\usepackage{tikz}
\usepackage{array,multirow,graphicx}
\usepackage{amsmath}
\usepackage{multirow}
\usepackage{footmisc}
\usepackage{moreverb}
\usepackage[utf8]{inputenc} 
\usepackage[algo2e,ruled,vlined,linesnumbered]{algorithm2e}
\usepackage{tikz}
\usepackage{graphicx}
\usepackage{pgfplots}
\pgfplotsset{compat=1.13}
\usepackage{algpseudocode}
\usepackage{array}
\usepackage{multirow}
\usepackage{tabularx}
\usepackage[none]{hyphenat}
\usepackage{amssymb}
\usepackage{colortbl}

\makeatother

\begin{document}%

%
\authorA{Fabio L\'opez-Pires}
\affA{Itaipu Technological Park, National University of Asunci\'on - Paraguay\\fabio.lopez@pti.org.py}

\authorB{Lino Chamorro}
\affB{National University of Asunci\'on - Paraguay\\lchamorro@pol.una.py}

\authorC{Benjam\'in Bar\'an}
\affC{National University of Asunci\'on - Paraguay\\bbaran@pol.una.py}

\hyphenation{pro-blems pro-po-sing con-si-de-ring accor-ding po-ssi-ble taxo-no-my stu-died o-pe-ra-tio-nal a-pproa-ches e-co-no-mi-cal a-ddre-ssed de-mons-tra-ted a-pproach pro-blem fo-llo-wing cha-rac-te-ris-tics con-so-li-da-ting exam-ple con-si-de-ra-tions summa-rizes se-ve-ral}

\setcounter{page}{1}

\LRH{Fabio L\'opez-Pires, Lino Chamorro and Benjam\'in Bar\'an}

\VOL{x}

\ISSUE{x}

\PUBYEAR{2020}

\BottomCatch


\PUBYEAR{2020}

\subtitle{}

\title{A Multi-Objective Approach for Multi-Cloud Infrastructure Brokering in Dynamic Markets}

\begin{abstract}
Cloud Service Brokers (CSBs) facilitate complex resource allocation decisions, efficiently mapping dynamic tenant demands onto dynamic provider offers, where several objectives should ideally be considered. This work proposes for the first time a pure multi-objective formulation of a broker-oriented Virtual Machine Placement (VMP) problem for dynamic environments, simultaneously optimizing the following objective functions: (i) Total Infrastructure CPU (TICPU), (ii) Total Infrastructure Memory (TIMEM) and (iii) Total Infrastructure Price (TIP) while considering load balancing across providers. To solve the formulated multi-objective problem, a Multi-Objective Evolutionary Algorithm (MOEA) is proposed. Considering that each time a demand (or offer) change occurs, a set of non-dominated solutions is found by Pareto-based algorithms as the one proposed, different selection strategies were evaluated in order to automatically select a convenient solution. Additionally, the proposed algorithm, including the considered selection strategies, was compared against mono-objective state-of-the-art alternatives in different scenarios with real data from providers in actual markets. Experimental results demonstrate that a pure multi-objective optimization approach considering the preferred solution selection strategy (S3) outperformed other mono-objective evaluated alternatives.
\end{abstract}

\KEYWORD{Virtual Machine Placement; Cloud Infrastructure; Multi-Objective Optimization; Dynamic Brokerage; Cloud Computing; Evolutionary Algorithm.}

\maketitle

\section{Introduction}
\label{sec:introduction}

Cloud computing datacenters deliver Infrastructure (IaaS), Platform (PaaS), and Software (SaaS) as services, available to end users in a pay-as-you-go basis. Particularly, in the IaaS model, customers can temporarily purchase processing, storage, network, and other fundamental computing resources to deploy arbitrary software, which can include different operating systems and applications \cite{buyya2008market}.

Cloud computing markets are currently composed by a wide ecosystem of heterogeneous Cloud Service Providers (CSPs) with different pricing schemes, virtual machine (VM) offers and features. Likewise, there are heterogeneous Cloud Service Tenants (CSTs) with different requirements and budgets to deploy complex cloud infrastructures. Cloud Service Brokers (CSBs) play a strategic role providing the CSTs with abstractions of complex resource allocation decisions, mapping specific requirements of each cloud infrastructure according to particular CST preferences (demands) onto available resources of CSPs (offers) \cite{LucasSimarro2011}.

Efficiently mapping demands from CSTs to available CSP offers could be defined as broker-oriented Virtual Machine Placement (VMP) \cite{lopez2015}. In the VMP literature, this process is also known as Cloud Application Brokerage (CAB) \cite{Tordsson2012}. 

CSBs allow CSTs to deploy multi-cloud infrastructures, i.e. an infrastructure composed by cloud resources allocated on different CSPs, avoiding vendor lock-in problems while optimizing infrastructure costs and improving performance \cite{Tordsson2012}.

Broker-oriented VMP problems could be studied in both static and dynamic environments \cite{lopez2015}. In static environments, CSP offers and CST requirements do not change over time; while in real-world dynamic environments CSPs could for example change VMs features, offers and prices over time, while CSTs could elastically increase or decrease their requirements for computational resources or change other parameters such as available budget. Considering these examples of dynamic environments, static approaches applied in current dynamic markets potentially result in sub-optimal solutions of broker-oriented VMP problems. Additionally, trending dynamic markets of Infrastructure as a Service (IaaS) present competitive advantages that could be exploited by formulating the broker-oriented VMP problem considering varying parameters in actual cloud markets, as dynamic environments. Consequently, proposing first formulations to broker-oriented VMP problems, according to particular requirements of dynamic environments, represent the main focus of the present research work.

Once a particular environment is defined for a broker-oriented VMP problem; it could be formulated considering both mono-objective or multi-objective approaches \cite{lopez2015}. Additionally, several objectives could be considered when studying a broker-oriented VMP problem, depending on the selected optimization approach (see Table \ref{tab:state-of-the-art}). Considering that several different objective functions were already studied in the broker-oriented VMP literature, these important objectives should be ideally taken into account simultaneously, as previously studied for static environments. Consequently, studying particular issues related to multi-objective optimization when proposing solution techniques to broker-oriented VMP problems in dynamic environments represent another main challenge addressed as part of this research work.

To the best of the authors’ knowledge, there is no published work presenting a multi-objective formulation of the considered problem for dynamic environments. Based on previous research on the field, the considered dynamic environment includes: (i) dynamic VM type offers from CSPs, (ii) dynamic pricing schemes from CSPs and (iii) variable number of required VMs from CSTs. Consequently, this work proposes for the first time a pure multi-objective formulation of a broker-oriented VMP problem in dynamic environments for the simultaneous optimization of the following three objective functions: (i) Total Infrastructure CPU (TICPU), (ii) Total Infrastructure Memory (TIMEM) and (iii) Total Infrastructure Price (TIP) while considering a Minimum Distribution Index ($LOC_{min}$) constraint to meet load balancing of VMs among CSPs, avoiding vendor lock-in problems. It should be noted that according to the studied literature (see Table \ref{tab:state-of-the-art}), TICPU and TIP are the most studied objective functions, resulting in the main selection criterion of objective functions for this work. Additionally, as a main resource for any type of virtual infrastructure or application, RAM Memory is also considered, denoted as TIMEM. These three considered objective functions were selected as a first possible formulation. By no means, the authors claim that the proposed formulation is the most appropriate one, mainly considering that each CST may have different preferences according to the nature of the required cloud infrastructure.

To solve the formulated multi-objective problem and to be able to effectively scale the resolution of the formulated broker-oriented VMP problem to large problem instances, a MOEA is proposed. Considering that the output of the proposed algorithm is a set of non-dominated solutions, a single solution could be manually selected by a decision maker according to particular needs. To facilitate this task, this work also evaluates different selection strategies to automatically select a convenient solution at each stage, i.e. at any change of the considered parameters that compose a dynamic environment. The proposed algorithm, including the considered selection strategies, was evaluated in different scenarios with real data from cloud computing providers against mono-objective state-of-the-art alternatives.

Considering that the most studied (hard) constraint represents load balancing of VMs across available CSPs to avoid vendor lock-in problems, it is also considered in this work, as well as (soft) optional constraints associated to each objective function cost to guide the decision space exploitation and reduce the number of non-dominated solutions in an obtained Pareto set approximation \cite{von2014survey}.

\hspace{1mm}

In summary, the main contributions of this work are:

\begin{itemize}
	\item a first pure multi-objective formulation of a broker-oriented VMP problem, simultaneously optimizing three objective functions: (i) TICPU, (ii) TIMEM and (iii) TIP, while considering load balancing across providers for dynamic environments including: (i) dynamic VM type offers from CSPs, (ii) dynamic pricing schemes from CSPs and (iii) variable number of required VMs from CSTs;
	\item a Multi-Objective Evolutionary Algorithm (MOEA) that is able to effectively solve large-scale instances of the proposed formulation of the problem; and
	\item an experimental evaluation of three strategies for automatically selecting a convenient solution from a Pareto set approximation for the studied problem, considering: (S1) random, (S2) minimum distance to origin, (S3) preferred solution, against mono-objective state-of-the-art alternatives (see Section \ref{sec2:selection_strategies}).
\end{itemize}
	
The remainder of this work is organized as follows: Section \ref{sec:related_work} presents a summary of the related work of the considered problem and the motivation of this research. Section \ref{sec:MOP} introduces general concepts of Pure Multi-Objective Optimization (PMO) problems. Next, Section \ref{sec:problem_formulation} summarizes a multi-objective formulation of a broker-oriented VMP problem, considering the simultaneous optimization of three objective functions for dynamic environments. Section \ref{sec:MOEA} presents the MOEA proposed for solving the formulated multi-objective problem, while Section \ref{sec:experimental_results} summarizes the experimental results and presents the main findings of this research work. Finally, conclusions and future work are left for Section \ref{sec:conclusions}.

\section{Related Works}
\label{sec:related_work}

Broker-oriented VMP problems have been mostly studied in static environments in the specialized literature, considering both mono-objective and multi-objective approaches for the optimization of several relevant objective functions mainly related to the capacity of resources of the requested cloud infrastructure and its total economical costs (see Table \ref{tab:state-of-the-art}).  Dynamic environments were briefly explored considering also several important objective functions, but only one at a time with a mono-objective optimization approach (see Table \ref{tab:state-of-the-art}). These existing objectives should be ideally studied considering a multi-objective optimization approach, simultaneously optimizing more than one objective function in trending dynamic markets. 

This work proposes a first multi-objective approach for a broker-oriented VMP problem in dynamic environments, considering the relevance of this type of brokerage technique for trending dynamic cloud computing markets, as previously described in Section \ref{sec:introduction}.

\begin{table}[!t]
	\centering
    \begin{tabular}{|c|c|c|c|c|}
      \hline
			\multirow{2}{*}{\textbf{Environment}} & \textbf{Optimization} & \multirow{2}{*}{\textbf{Objective Function(s)}} & \textbf{Solution} & \multirow{2}{*}{\textbf{Reference}} 			\\
																						 & \textbf{Approach}		 &																							 & \textbf{Technique} &				\\
			\hline
			&&&& \\
			\multirow{14}{*}{Static}& \multirow{6}{*}{Mono-Objective}  & Total Infrastructure CPU (TICPU)  & ILP & 	\cite{Tordsson2012} \\
			&&&& \\															
			\cline{3-5}
			&&&& \\
															&																	 & \multirow{4}{*}{Total Infrastructure Price (TIP)} 		& SIP & \cite{chaisiri2009}  \\
			&&&& \\															
			\cline{4-5}															
			&&&& \\
															&																	 & 																											& Hybrid EA & \cite{Mark2011}\\ 
			&&&& \\												
			\cline{2-5}
			&&&& \\
															& \multirow{8}{*}{Multi-Objective}& CST's Satisfaction 									& \multirow{2}{*}{MOEA} &\multirow{2}{*}{\cite{Kessaci2013}}  			\\
															&																   & and CSB's Profit 								 	&	&	 																	 		\\ 
			&&&& \\												
			\cline{3-5}
			&&&& \\
															&																	 & Total Infrastructure Price (TIP)		& \multirow{4}{*}{Rules Engine} & \multirow{4}{*}{\cite{Amato2013,Amato2013a}}\\
															&																	 & Total Infrastructure CPU (TICPU) 		& &																		   \\
															&																	 & CSP's Reputation											&	&																		   \\															 &																	& and Availability	 					 				 & &		 															    \\			
			&&&& \\
			\hline
			&&&& \\
			\multirow{12}{*}{Dynamic} & \multirow{6}{*}{Mono-Objective} & Total Infrastructure Price (TIP)		&	ILP & \cite{LucasSimarro2011} 				\\
			&&&& \\
			\cline{3-5}
			&&&& \\
															 &																 & Total Infrastructure Capacity (TIC)	& ILP & \cite{Li2011} 									\\
															 &																 & Total Infrastructure Price (TIP)			&	&																			\\	
			\cline{4-5}	
															 &																 & or Migration Cost (MC)								&	&																			\\
																&																	&																			&	GA &	\cite{chamorro2016}							\\
			&&&& \\
			\cline{2-5}
			&&&& \\
															 & \multirow{3}{*}{\textit{Multi-Objective}}	& \textit{Total Infrastructure CPU (TICPU)}	& \multirow{3}{*}{\textit{MOEA}} &\multirow{3}{*}{\textit{This work}}						\\
															 &																	& \textit{Total Infrastructure Memory (TIMEM)}	&	&																			\\
															 &																	& \textit{Total Infrastructure Price (TIP)} 		&	& 																		\\																								
			&&&& \\
			\hline												
	\end{tabular}
\caption{Summary of related works studying broker-oriented VMP problems.}
\label{tab:state-of-the-art}
\end{table}

\subsection{Mono-Objective Brokerage in Static Environments}

Tordsson et al. proposed in \cite{Tordsson2012}, scheduling algorithms for the CAB problem in static environments, taking into account fixed CST requirements, fixed possible number of VM hardware configurations and known hourly prices for running VMs in a CSP. The proposed Integer Linear Programming (ILP) model considers the maximization of the Total Infrastructure Capacity (TIC) while defining a maximum budget constraint \cite{Tordsson2012}. CSTs may also consider the following deployment constraints: (i) VM hardware configurations, where a minimum and maximum instance type indexes are requested, (ii) number of VMs of each instance type, where a minimum and maximum number of VMs of each instance type are specified and (iii) load balancing, where a minimum and maximum percentage of VMs that can be located at each CSP are defined. It should be mentioned that the TIC objective proposed in \cite{Tordsson2012} represents a particular capacity of CPU rather than other possible resources. Consequently, to avoid ambiguous terminology, this work considers the mentioned objective function as Total Infrastructure CPU (TICPU).
 
Chaisiri et al. proposed in \cite{chaisiri2009}, a Stochastic Linear Programming (SIP) model to minimize the costs associated to hosting VMs in a multi-cloud deployment architecture under future demand and price uncertainty, subject to constraints to ensure that the requested demand is met and the allocations of VMs do not exceed the resource capacity offered by CSPs. This work considers the costs minimized in \cite{chaisiri2009} as Total Infrastructure Price (TIP). The proposed model considers two payment plans for CSTs: (i) reservation and (ii) on-demand, as offered by Amazon Web Services \cite{amazon2015}. Extending the work proposed in \cite{chaisiri2009}, Mark et al. proposed in \cite{Mark2011} an Evolutionary Optimal Virtual Machine Placement (EOVMP) algorithm with a demand forecaster to allocate VMs using reservation and on-demand plans for job processing. The proposed EOVMP is a hybridized algorithm of Genetic Algorithm (GA), Particle Swarm Optimization (PSO) and Ant Colony Optimization (ACO). The model proposed in \cite{Mark2011} minimizes the total price of the virtual infrastructure, denoted in this work as TIP. In the first stage, the proposed demand forecaster predicts the demand and next, the EOVMP considers the predicted demand to allocate the necessary VMs using both reservation and on-demand plans. Constraints are related to provisioning phases in order to meet CST demands.

\subsection{Multi-Objective Brokerage in Static Environments}

Kessaci et al. proposed in \cite{Kessaci2013}, a job scheduler using a Multi-Objective Evolutionary Algorithm (MOEA) for response time and service price optimization in order to maximize the CST's satisfaction and simultaneously maximize the CSB's profit, providing a Pareto set of non-dominated solutions rather than a single solution.

Amato and Venticinque defined in \cite{Amato2013,Amato2013a} that cloud brokering problems may need to deal with several and, at the same time, contradictory objectives, finally combining all objectives into just one objective function with a weighted sum method. The considered objectives include \cite{Amato2013,Amato2013a}: (i) TIP, (ii) TICPU, (iii) CSP's Reputation and/or (iv) Availability. The mentioned objective functions could also be considered as two different types of constraints: mandatory (hard) and optional (soft) constraints. The developed brokering tool is flexible enough to define a custom optimization model to meet particular CST requirements \cite{Amato2013,Amato2013a}.

\subsection{Mono-Objective Brokerage in Dynamic Environments}

Lucas-Simarro et al. proposed in \cite{LucasSimarro2011}, an optimization model for CSTs virtual cluster placement across available CSPs offers. This scheduler considers average prices or cloud price trends to suggest an optimal multi-cloud deployment. A mono-objective approach is considered for TIP minimization, selecting the best possible combination of CSPs that offer the lowest prices considering an ILP formulation. The following constraints are also considered in the proposed model \cite{LucasSimarro2011}: (i) distance, representing a minimum and maximum number of VMs that can be relocated across CSPs (to guarantee a certain number of VMs working all the time) and (ii) load balancing, where a minimum and maximum percentage of all VMs should be located at each CSP (to avoid vendor lock-in problems).

Li et al. proposed in \cite{Li2011}, an ILP formulation for cloud service brokering in dynamic environments, where instance types, prices and service performance are continuously changing throughout the service life-cycle. The mentioned work proposed three different mono-objective optimization models: (i) TIC maximization, (ii) TIP minimization and (iii) Migration Costs (MC) minimization. The following constraints are applied to each optimization model: (i) budget, where TIP cannot exceed a specified budget limit, (ii) unique placement, where each VM has to be of exactly one instance type and placed in exactly one CSP and (iii) load balancing, representing a minimum and maximum percentage of all VMs to be located at each CSP. These mentioned optimization models were experimentally evaluated considering different dynamic scenarios: (i) new instance type offers, (ii) changing prices and (iii) service performance elasticity (number of VMs).

Considering the scalability limitations of the ILP formulation proposed in \cite{Li2011}, Chamorro et al. presented in \cite{chamorro2016} a Genetic Algorithm (GA) for dynamic cloud application brokerage also considering a mono-objective optimization approach. The mentioned work studies a broker-oriented VMP problem with large problem instances considering dynamic environments such as: (i) variable resource offers and (ii) varying pricing from CSPs, as well as (iii) dynamic requirements of CSTs.

\section{Pure Multi-Objective Optimization (PMO)}
\label{sec:MOP}
A general Pure Multi-Objective Optimization (PMO) problem includes a set of $p$ decision variables, $q$ objective functions, and $r$ constraints. Objective functions and constraints are functions of decision variables. In a PMO formulation, $x$ represents the decision vector, while $y$ represents the objective vector. The decision space is denoted by $X$ and the corresponding objective space as $Y$. These can be expressed as \cite{coello2007}:

\vspace{1mm}

$Optimize:$

\vspace{-2mm}

\begin{equation} y = f(x) = [f_1(x), f_2(x), \ldots, f_q(x)]
\label{eq:opti}
\end{equation}

\vspace{-2mm}

\textit{subject to:}

\vspace{-2mm}

\begin{equation} e(x) = [e_1(x), e_2(x), \ldots, e_r(x)] \geq 0
\label{eq:subj}
\end{equation}

\vspace{-2mm}

$where:$

\vspace{-2mm}

\begin{equation} x = [x_1, x_2, \ldots, x_p] \in X
\label{eq:set1}
\end{equation}

\vspace{-2mm}

\begin{equation} y = [y_1, y_2, \ldots, y_p] \in Y
\label{eq:set2}
\end{equation}

Note that optimizing can mean maximizing or minimizing, depending on the particular problem context. A set of constraints $e(x) \geq 0$ defines a set of feasible solutions $X_f \subset X$ and its corresponding set of feasible objective vectors $Y_f \subset Y$. The feasible decision space $X_f$ is the set of all decision vectors $x$ in a decision space $X$ that satisfy constraints $e(x)$, and it is defined as:

\vspace{-2mm}

\begin{equation} X_f = \left\{x|x \in X \wedge e(x) \geq 0 \right\}
\label{eq:max1}
\end{equation}

The feasible objective space $Y_f$ is the set of the objective vectors $y$ that represents the image of $X_f$ onto $Y$ and it is denoted by:

\vspace{-2mm}

\begin{equation} Y_f = \left\{y|y \in f(x) \quad \forall x \in X_f \right\}
\label{eq:max2}
\end{equation}

To compare two solutions in a pure multi-objective context, the concept of Pareto dominance is used. Given two feasible solutions $u$, $v \in X$, $u$ dominates $v$, denoted as $u \succ v$, if $f(u)$ is better or equal to $f(v)$ in every objective function and strictly better in at least one objective function. If neither $u$ dominates $v$, nor $v$ dominates $u$, $u$ and $v$ are said to be non-comparable (denoted as $u \sim v$).

A decision vector $x$ is non-dominated with respect to a set $U$, if there is no member of $U$ that dominates $x$. The set of non-dominated solutions of the whole set of feasible solutions $X_f$, is known as the optimal Pareto set $P^*$. The corresponding set of objective vectors constitutes the optimal Pareto front $PF^*$.

\section{Proposed Multi-Objective Broker-oriented VMP Formulation}
\label{sec:problem_formulation}

\begin{figure}[!t]
	\centering
	\includegraphics[scale=0.5]{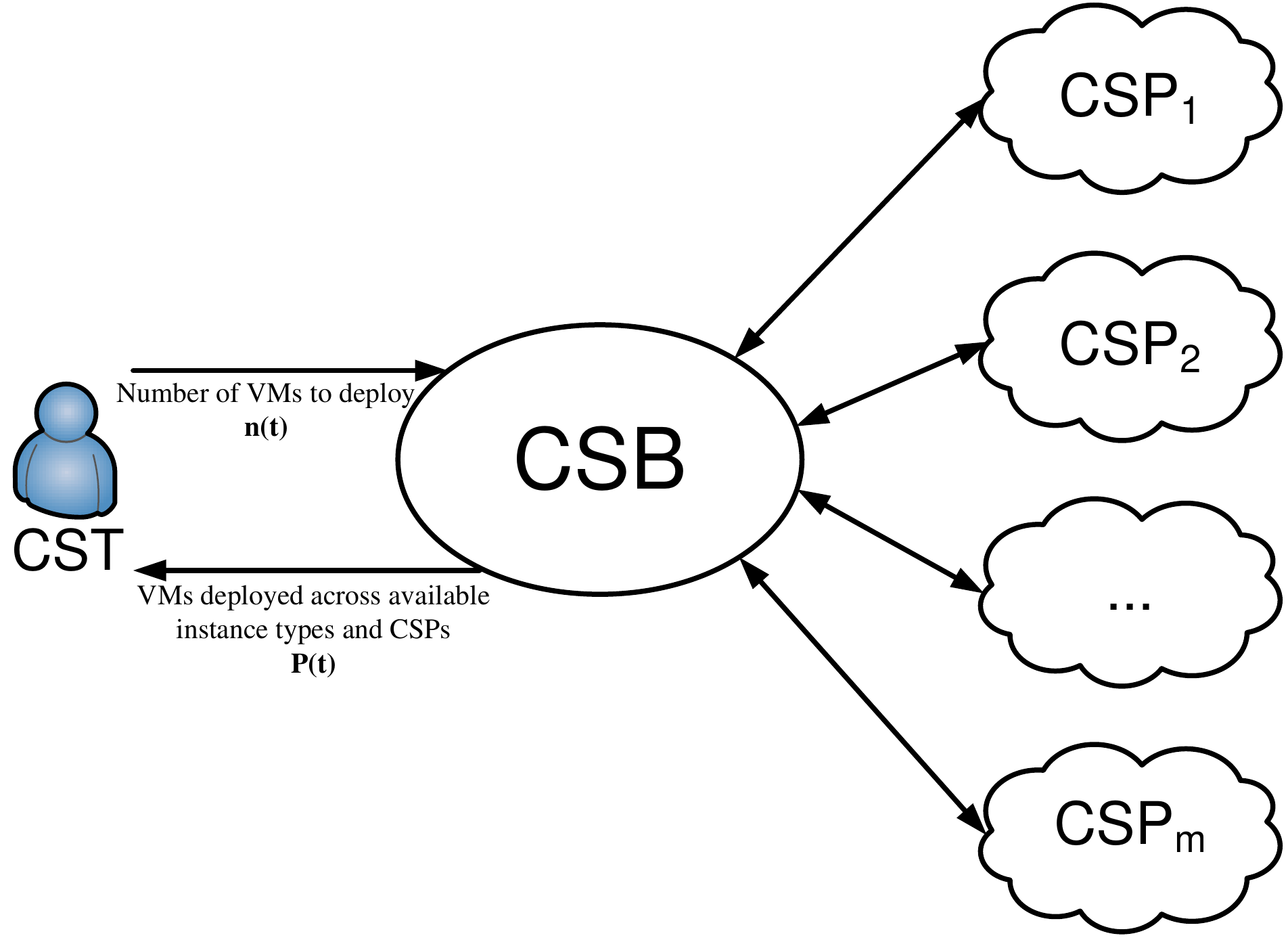}
	\caption{A broker-oriented VMP problem considering dynamic instance types and price offers from CSPs as well as dynamic demands from CSTs.}\label{fig:vmp}
\end{figure}

To the best of the authors’ knowledge this work presents for the first time a formulation of a broker-oriented VMP problem for dynamic environments, considering this time the simultaneous optimization of the following three objective functions: (i) TICPU, (ii) TIMEM and (iii) TIP while considering load balancing of VMs among CSPs. Formally, the proposed pure multi-objective broker-oriented VMP problem formulation can be enunciated as follows (see Figure \ref{fig:vmp}). 

Given: 

\begin{itemize}
	\item a set of $m$ CSPs; 
	\item a set of $l(t)$ instance types $IT_j$ available at each CSP $c_{k}$ (denoted as $IT_{j,k}$), including its characteristics (as explained in Section \ref{sec:problem_inputs});
	\item a set of $n(t)$ VMs $v_i$ to be deployed across available CSPs $c_{k}$, including an expected lifetime of the requested infrastructure;
\end{itemize}

\noindent it is sought convenient combinations of instance types and CSPs to deploy the requested VMs in a dynamic cloud computing market, satisfying the constraints of the problem while simultaneously optimizing all objective functions defined in this formulation in a pure multi-objective context, before selecting a specific solution at each instant $t$. Based on hourly prices, each instant $t$ typically represents an hour of resource provisioning and placement reconfiguration is triggered for next instant $t+1$ only if a change on any of the dynamic parameters considered for this particular dynamic cloud computing market is detected (see Figure \ref{fig:vmp}).

\subsection{Problem Inputs}
\label{sec:problem_inputs}

The proposed broker-oriented VMP problem receives information as input data, maintaining an updated information on CSP offers and CST requests. Thus, this work considers a dynamic environment composed by the following dynamic parameters: (i) new instance types may be introduced in the cloud market, (ii) changes in prices of resources and (iii) changes in the number of requested VMs.

The set of available CPU resources in the cloud computing market is represented as a matrix $CPU(t) \in \mathbb{Z}^{l(t) \times m}$, where each element $CPU_{j,k}$ represents the available CPU resources associated to each instance type $IT_j$ at each $c_k$ CSP.

\begin{equation}
\label{eq:matrix_CPU}
\hspace{1cm}
  CPU(t) =
  \left[ 
  \begin{array}{cccc}
  CPU_{1,1} & CPU_{1,2} & \dots & CPU_{1,m} \\
  \dots  & \dots  & \dots  & \dots  \\
  CPU_{l(t),1} & CPU_{l(t),2} & \dots & CPU_{l(t),m} \\
  \end{array}
  \right] 
\end{equation}
where:

\vspace{1.5mm}

\noindent \begin{tabular}{p{1.3cm} p{10.2cm}}
$CPU_{j,k}$:	& CPU resources [in \# of cores] associated to each $IT_j$ at each $c_k$; \\
$l(t)$: & Number of instance types $IT_j$ at time instant $t$, where $1 \leq j \leq l(t)$;	\\
$m$: 		&	Number of CSPs $c_k$, where $1 \leq k \leq m$.	\\
\end{tabular}

\vspace{2mm}

Similarly, the set of available RAM memory resources is represented as a matrix $MEM(t) \in \mathbb{Z}^{l(t) \times m}$. This set of resources can be formulated as:

\begin{equation}
\label{eq:matrix_MEM}
\hspace{0.5cm}
  MEM(t) =
  \left[ 
  \begin{array}{cccc}
  MEM_{1,1} & MEM_{1,2} & \dots & MEM_{1,m} \\
  \dots  & \dots  & \dots  & \dots  \\
  MEM_{l(t),1} & MEM_{l(t),2} & \dots & MEM_{l(t),m} \\
  \end{array}
  \right] 
\end{equation}
where:

\vspace{1.5mm}

\noindent \begin{tabular}{p{1.3cm} p{10.2cm}}
$MEM_{j,k}$:	& Memory resources [in GB] associated to each $IT_j$ at each $c_k$; \\
$l(t)$: & Number of instance types $IT_j$ at time instant $t$, where $1 \leq j \leq l(t)$;	\\
$m$: 		&	Number of CSPs $c_k$, where $1 \leq k \leq m$.	\\
\end{tabular}

\vspace{2mm}

It is important to note that matrices $CPU(t)$ and $MEM(t)$ are functions of time because new instance type offers may be introduced in the cloud computing market (e.g. Amazon EC2 announced micro instances \cite{amazon2009,amazon2014}). Consequently, the number of instance types $l(t)$ may vary, representing potential placement reconfigurations.

\vspace{1.5mm}

The set of prices associated to each $IT_j$ at each CSP $c_{k}$ is represented as a matrix $PRC(t) \in \mathbb{R}^{l(t) \times m}$, where each element $PRC_{j,k}(t)$ represents its hourly price.

\begin{equation}
\label{eq:matrix_PRC}
\hspace{0.4cm}
  PRC(t) =
  \left[ 
  \begin{array}{cccc}
  PRC_{1,1}(t) & PRC_{1,2}(t) & \dots & PRC_{1,m}(t) \\
  \dots  & \dots  & \dots  & \dots  \\
  PRC_{l(t),1}(t) & PRC_{l(t),2}(t) & \dots & PRC_{l(t),m}(t) \\
  \end{array}
  \right] 
\end{equation}
where:

\vspace{1.5mm}

\noindent \begin{tabular}{p{1.5cm} p{10cm}}
$PRC_{j,k}(t)$:	    & Price [in \$] associated to each $IT_j$ at each $c_k$; \\
$l(t)$: & Number of instance types $IT_j$ at time instant $t$, where $1 \leq j \leq l(t)$;	\\
$m$: 		&	Number of CSPs $c_k$, where $1 \leq k \leq m$.	\\
\end{tabular}

\vspace{2mm}

It should be noted that dynamic pricing schemes are considered, covering different possible schemes such as spot prices \cite{amazon2015} or even special discounts \cite{amazon2016}. In this context, prices $PRC_{j,k}(t)$ may change over time, where economical opportunities could be exploited by conveniently reconfiguring a current placement.

Additionally, to consider an overhead for placement reconfigurations in the proposed simulation, information about allocation $(AT_{j,k})$ and release times $(RT_{j,k})$ [in hours] of each $IT_{j,k}$ should also represented in the input data of the formulated broker-oriented VMP problem. 

These information could be experimentally obtained by CSBs, as presented by Iosup et al. in \cite{iosup2011performance}. Similarly to the above presented matrices, both sets of allocation $(AT_{j,k})$ and release $(RT_{j,k})$ times associated to each $IT_{j,k}$ are represented as matrices $AT(t)$ and $RT(t) \in \mathbb{R}^{l(t) \times m}$ respectively. These can be formulated as:

\begin{equation}
\label{eq:matrix_AT}
\hspace{2cm}
  AT(t) =
  \left[ 
  \begin{array}{cccc}
  AT_{1,1} & AT_{1,2} & \dots & AT_{1,m} \\
  \dots  & \dots  & \dots  & \dots  \\
  AT_{l(t),1} & AT_{l(t),2} & \dots & AT_{l(t),m} \\
  \end{array}
  \right] 
\end{equation}

and:

\begin{equation}
\label{eq:matrix_RT}
\hspace{2cm}
  RT(t) =
  \left[ 
  \begin{array}{cccc}
  RT_{1,1} & RT_{1,2} & \dots & RT_{1,m} \\
  \dots  & \dots  & \dots  & \dots  \\
  RT_{l(t),1} & RT_{l(t),2} & \dots & RT_{l(t),m} \\
  \end{array}
  \right] 
\end{equation}
where:

\vspace{1.5mm}

\noindent \begin{tabular}{p{1.3cm} p{10.2cm}}
$AT_{j,k}$:	& Allocation time [in hours] associated to each $IT_j$ at each $c_k$; \\
$RT_{j,k}$:	& Release time [in hours] associated to each $IT_j$ at each $c_k$; \\
$l(t)$: 		& Number of instance types $IT_j$ at time instant $t$, where $1 \leq j \leq l(t)$;	\\
$m$: 				&	Number of CSPs $c_k$, where $1 \leq k \leq m$.	\\
\end{tabular}

\vspace{2mm}

Finally, CSTs must specify the number of requested VMs to be deployed, denoted as $n(t)$. It is important to consider that the number of VMs could be dynamically adjusted according to particular CST requirements. Additionally, an estimated lifetime of the infrastructure [in hours], denoted as $H(t)$, is provided by CSTs in order to calculate if a placement reconfiguration is convenient, taking into account the time that the requested VMs will remain in operation.

\subsection{Problem Outputs}
\label{sec:problem_outputs}

A matrix $P(t) \in \mathbb{B}^{n(t) \times l(t) \times m}$, composed by decision variables $x_{i,j,k}(t)$, represents a possible instance type selection and placement of VMs on available CSPs at instant $t$. Consequently, the output data of the proposed multi-objective broker-oriented VMP problem is a new matrix $P(t+1) \in \mathbb{B}^{n(t+1) \times l(t+1) \times m}$, composed by $x_{i,j,k}(t+1)$, representing a new instance type selection and placement of VMs on available CSPs at the next instant $t+1$, considering changes presented in the cloud market. It is important to note that $P(t+1)$ is selected in this work from a set of non-dominated solutions (Pareto set approximation).

\subsection{Constraints}
\label{sec:constraints}

Feasible solutions of the proposed broker-oriented VMP problem is restricted by the following constraints:

\begin{itemize}
	\item unique placement of VMs, see Section \ref{sec:unique_placement};
	\item load balancing of VMs among CSPs, see Section \ref{sec:load_balancing};
	\item lower and upper bounds associated to each objective function $f_1(x)$ to $f_3(x)$, see Section \ref{sec:soft_constraints}.
\end{itemize}

Each of these constraints is mathematically defined in the following sub-sections.

\subsubsection{Unique placement of VMs}
\label{sec:unique_placement}

Each VM $v_i$, $\forall i \in [1, \dots, n(t)]$, should be provisioned selecting a single instance type $IT_j$, $\forall j \in [1, \dots, l(t)]$, and located to run on a single CSP $c_k$, $\forall k \in [1, \dots ,m]$. Consequently, this unique placement constraint is mathematically expressed as:

\begin{equation} 
\sum_{j=1}^{l} \sum_{k=1}^{m} x_{ijk} = 1, \forall i \in [1,...,n]
\label{eq2:unique_placement} 
\end{equation}
where:

\vspace{1.5mm}

\noindent \begin{tabular}{p{1.8cm} p{15.0cm}}
$n(t)$: 			&	Number of requested VMs $v_i$, where $1 \leq i \leq n(t)$;	\\
$l(t)$: 			& Number of instance types $IT_j$ at time instant $t$, where $1 \leq j \leq l(t)$;	\\
$m$: 				&	Number of CSPs $c_k$, where $1 \leq k \leq m$;	\\
$x_{i,j,k}(t)$  :	& Binary variable equals 1 if $v_i$ is of instance type $IT_j$ and is located at CSP \\
                    & $c_k$ at instant $t$; otherwise, it is 0.	
\end{tabular}

\vspace{2mm}

\subsubsection{Load balancing of VMs among CSPs}
\label{sec:load_balancing}

As an alternative to avoid vendor lock-in problems \cite{Li2011}, a load balancing constraint $(LOC_{min})$ is modeled as a minimum percentage of VMs to be located at each CSP $c_k$, $\forall k \in [1, \dots, m]$. Consequently, this constraint is mathematically expressed as:

\begin{equation}
\hspace{1cm} 
LOC_{min} \leq \frac{\sum_{i=1}^{n(t)} \sum_{j=1}^{l(t)} x_{i,j,k}(t)}{n(t)} \quad \quad \forall k \in [1, \dots, m]
\label{eq:loc_min}
\end{equation}
where:

\vspace{1.5mm}

\noindent \begin{tabular}{p{1.3cm} p{15cm}}
$LOC_{min}$:	    & Minimum percentage of VMs to be located at each CSP $c_k$;  \\
$n(t)$: 			& Number of requested VMs $v_i$, where $1 \leq i \leq n(t)$;	\\
$l(t)$: 			& Number of instance types $IT_j$ at time instant $t$, where $1 \leq j \leq l(t)$;	\\
$m$: 				& Number of CSPs $c_k$, where $1 \leq k \leq m$;	\\
$x_{i,j,k}(t)$:	    & Binary variable equals 1 if $v_i$ is of instance type $IT_j$ and is located at CSP \\
                    & $c_k$ at instant $t$; otherwise, it is 0.
\end{tabular}

\vspace{2mm}

\subsubsection{Adjustable constraints}
\label{sec:soft_constraints}

Considering the multi-objective formulation of the proposed broker-oriented VMP problem to be summarized in Section \ref{sec:multi-objective_problem_formulation}, the set of non-dominated solutions may include a large number of feasible solutions, being increasingly difficult to discriminate among solutions using only the dominance relation \cite{flopezmany2015}. For this reason, this work proposes the utilization of lower and upper bounds associated to each objective function $z \in \{1, \dots, q\}$ $(L_z \leq f_z(x) \leq U_z)$ to be able to reduce the number of possible compromise solutions according to the CST particular requirements. 

Taking into account that CSTs may consider difficult to define appropriate bounds because these values are unknown a-priori, these lower and upper bounds are modeled as soft constraints \cite{Amato2013}, where a percentage of the expected bounds could be exceeded (e.g. if a CST expects at least 100 [GB] for TIMEM with an acceptable margin of 10\%, the cost of feasible solutions for $f_2(x)$ must be higher or equal to $L_2$=90 [GB]). Similarly, bounds could be adjusted for maximum expected values.

\subsection{Objective Functions}
\label{sec:objective_functions}

When one or more of the dynamic parameters considered in the proposed formulation present a change in the cloud computing market (see Section \ref{sec:problem_inputs}), new opportunities may be exploited by reconfiguring the current placement of requested VMs (e.g. migrating VMs among CSPs and/or changing selected instance types). Inspired in \cite{Li2011}, this work presents a Reconfiguration Overhead (RO) based on the wasted resources during each placement reconfiguration period, considered for both TICPU and TIMEM objectives to be presented in the following subsections.

\subsubsection{Total Infrastructure CPU}
\label{sec:TICPU}
For CPU resources, the $RO_{CPU}$ is modeled as the CPU capacity wasted during the reconfiguration period (i.e. allocation and release time [in hours] of modified VMs). This overhead can mathematically be formulated as:

\begin{equation}
RO_{CPU} = \sum_{i=1}^{n(t)} \sum_{\substack{j=1\\j\prime=1}}^{l(t)} \sum_{\substack{k=1\\k\prime=1}}^{m}[CPU_{j,k} \times RT_{j,k} \times x_{i,j,k}(t) + CPU_{j\prime,k\prime} \times AT_{j\prime,k\prime} \times x_{i,j\prime,k\prime}(t+1)] \\
\forall j \neq j\prime ; k \neq k\prime 
\label{eq:mc_cpu}
\end{equation}
where:

\vspace{1.5mm}

\noindent \begin{tabular}{p{1.3cm} p{15cm}}

$RO_{CPU}$: 	& Total CPU reconfiguration overhead;	\\
$n(t)$: 			&	Number of requested VMs $v_i$, where $1 \leq i \leq n(t)$;	\\
$l(t)$: 			& Number of instance types $IT_j$ at time instant $t$, where $1 \leq j \leq l(t)$;	\\
$m$: 					&	Number of CSPs $c_k$, where $1 \leq k \leq m$;	\\
$CPU_{j,k}$:	& CPU resources [in \# of cores] associated to each $IT_j$ at each $c_k$; \\
$RT_{j,k}$:		& Release time [in hours] associated to each $IT_j$ at each $c_k$; \\
$AT_{j,k}$:		& Allocation time [in hours] associated to each $IT_j$ at each $c_k$; \\
$x_{i,j,k}(t)$:	& Binary variable equals 1 if $v_i$ is of instance type $IT_j$ and is located at CSP \\
                & $c_k$ at instant $t$; otherwise, it is 0.
\end{tabular}

\vspace{2mm}

Additionally, the TICPU can be represented as:

\vspace{-3.5mm}

\begin{equation} 
\hspace{1.5cm} 
TICPU = H(t) \times \sum_{i=1}^{n(t)} \sum_{j=1}^{l(t)} \sum_{k=1}^{m} CPU_{j,k} \times x_{i,j,k}(t+1)
\label{eq:TICPU}
\end{equation}
where:

\vspace{1.5mm}

\noindent \begin{tabular}{p{1.3cm} p{15cm}}
$TICPU$: 					& Total Infrastructure CPU;	\\
$H(t)$:						& Expected remaining lifetime of the infrastructure [in hours];	\\
$l(t)$: 					& Number of instance types $IT_j$ at time instant $t$, where $1 \leq j \leq l(t)$;	\\
$m$: 							&	Number of CSPs $c_k$, where $1 \leq k \leq m$;	\\
$CPU_{j,k}$: 			& CPU capacity [in \# of cores] of instance type $IT_j$ at CSP $c_k$;	\\
$x_{i,j,k}(t)$:	    & Binary variable equals 1 if $v_i$ is of instance type $IT_j$ and is located at CSP \\
                    & $c_k$ at instant $t$; otherwise, it is 0.
\end{tabular}

\vspace{2mm}

Finally, the objective function $f_1(x)$ is expressed as the difference between the TICPU and its corresponding $RO_{CPU}$:

\vspace{-3.5mm}

\begin{equation} 
\hspace{3cm} 
f_1(x) = TICPU - RO_{CPU}
\label{eq:f_1(x)}
\end{equation}

\subsubsection{Total Infrastructure Memory}
\label{sec:TIMEM}

For RAM memory resources, the $RO_{MEM}$ is modeled as the RAM memory capacity wasted during the reconfiguration period (i.e. allocation and release time [in hours] of modified VMs). This overhead can be mathematically formulated as:

\vspace{-3.5mm}

\begin{equation}
RO_{MEM} = \sum_{i=1}^{n(t)} \sum_{\substack{j=1\\j\prime=1}}^{l(t)} \sum_{\substack{k=1\\k\prime=1}}^{m}(MEM_{j,k} \times RT_{j,k} \times x_{i,j,k}(t) + MEM_{j\prime,k\prime} \times AT_{j\prime,k\prime} \times x_{i,j\prime,k\prime}(t+1)) \\
\forall j \neq j\prime ; k \neq k\prime 
\label{eq:mc_mem}
\end{equation}
where:

\vspace{1.5mm}

\noindent \begin{tabular}{p{1.3cm} p{15cm}}
$RO_{MEM}$: 	& Total RAM memory reconfiguration overhead;	\\
$n(t)$: 			&	Number of requested VMs $v_i$, where $1 \leq i \leq n(t)$;	\\
$l(t)$: 			& Number of instance types $IT_j$ at time instant $t$, where $1 \leq j \leq l(t)$;	\\
$m$: 					&	Number of CSPs $c_k$, where $1 \leq k \leq m$;	\\
$MEM_{j,k}$:	& RAM memory resources [in GB] associated to each $IT_j$ at each $c_k$; \\
$RT_{j,k}$:		& Release time [in hours] associated to each $IT_j$ at each $c_k$; \\
$AT_{j,k}$:		& Allocation time [in hours] associated to each $IT_j$ at each $c_k$; \\
$x_{i,j,k}(t)$:	    & Binary variable equals 1 if $v_i$ is of instance type $IT_j$ and is located at CSP \\
                    & $c_k$ at instant $t$; otherwise, it is 0.
\end{tabular}

\vspace{2mm}

Additionally, the TIMEM can be represented as:

\vspace{-3.5mm}

\begin{equation} 
\hspace{1.5cm} 
TIMEM = H(t) \times \sum_{i=1}^{n(t)} \sum_{j=1}^{l(t)} \sum_{k=1}^{m} MEM_{j,k} \times x_{i,j,k}(t+1)
\label{eq:TIMEM}
\end{equation}
where:

\vspace{1.5mm}

\noindent \begin{tabular}{p{1.3cm} p{15cm}}
$TIMEM$: 			& Total Infrastructure RAM memory;	\\
$H(t)$:				& Expected remaining lifetime of the infrastructure [in hours];	\\
$n(t)$: 			&	Number of requested VMs $v_i$, where $1 \leq i \leq n(t)$;	\\
$l(t)$: 			& Number of instance types $IT_j$ at time instant $t$, where $1 \leq j \leq l(t)$;	\\
$m$: 					&	Number of CSPs $c_k$, where $1 \leq k \leq m$;	\\
$MEM_{j,k}$: 	& RAM memory capacity [in GB] of instance type $IT_j$ at CSP $c_k$;	\\
$x_{i,j,k}(t)$:	    & Binary variable equals 1 if $v_i$ is of instance type $IT_j$ and is located at CSP \\
                    & $c_k$ at instant $t$; otherwise, it is 0.
\end{tabular}

\vspace{2mm}

Finally, the objective function $f_2(x)$ is expressed as the difference between the TIMEM and its corresponding $RO_{MEM}$:

\vspace{-3.5mm}

\begin{equation} 
\hspace{3cm} 
f_2(x) = TIMEM - RO_{MEM}
\label{eq:f_2(x)}
\end{equation}

\subsubsection{Total Infrastructure Price}
\label{sec:TIP}

The TIP of the requested infrastructure can be mathematically formulated as:

\begin{equation}
\hspace{0.5cm} 
f_3(x) = TIP = H(t) \times \sum_{i=1}^{n(t)} \sum_{j=1}^{l(t)} \sum_{k=1}^{m} PRC_{j,k}(t) \times x_{i,j,k}(t+1)
\label{eq:TIP}
\end{equation}
where:

\vspace{1.5mm}

\noindent \begin{tabular}{p{1.5cm} p{12cm}}
$TIP$: 						& Total Infrastructure Price;	\\
$H(t)$: 					& Expected lifetime of the infrastructure [in hours];	\\
$n(t)$: 					&	Number of requested VMs $v_i$, where $1 \leq i \leq n(t)$;	\\
$l(t)$: 					& Number of instance types $IT_j$ at time instant $t$, where $1 \leq j \leq l(t)$;	\\
$m$: 							&	Number of CSPs $c_k$, where $1 \leq k \leq m$;	\\
$PRC_{j,k}(t)$: 		& Price [in \$] of instance type $IT_j$ at CSP $c_k$;	\\
$x_{i,j,k}(t)$:	    & Binary variable equals 1 if $v_i$ is of instance type $IT_j$ and is located at CSP \\
                    & $c_k$ at instant $t$; otherwise, it is 0.
\end{tabular}

\subsection{Multi-Objective Problem Formulation}
\label{sec:multi-objective_problem_formulation}

In summary, a broker-oriented VMP problem based on constraints and multiple objectives previously detailed in Sections \ref{sec:constraints} and \ref{sec:objective_functions} respectively, may be written as:

\newpage

\noindent \textit{Maximize:}
\begin{equation}
\label{eq2:f(x)}
y_1 = f(x) = [f_1(x), f_2(x)] 
\end{equation}

\noindent and \textit{Minimize:}
\begin{equation}
\label{eq2:f(x1)}
y_2 = f(x) = [f_3(x)] 
\end{equation}

\noindent \textit{where:}

\begin{equation}
\label{eq:FOs}
\begin{aligned}
f_1(x) &= \text{Total Infrastructure CPU (TICPU);} 				\\
f_2(x) &= \text{Total Infrastructure Memory (TIMEM);} 		\\
f_3(x) &= \text{Total Infrastructure Price (TIP).} 				\\
\end{aligned}
\end{equation}

\noindent \textit{subject to:}

\begin{equation}
\label{eq2:constraints}
\begin{aligned}
	e_1(x) &: \text{unique placement of VMs;} 							\\
	e_2(x) &: \text{load balancing of VMs between CSPs;} 		\\
	e_3(x) &: f_1(x) \geq L_1; \\
	e_4(x) &: f_2(x) \geq L_2; \\
	e_5(x) &: f_3(x) \leq U_3; \\
\end{aligned}
\end{equation}

\section{Proposed Multi-Objective Evolutionary Algorithm (MOEA)}
\label{sec:MOEA}

In order to solve the proposed multi-objective broker-oriented VMP problem presented in Section \ref{sec:problem_formulation}, a Multi-Objective Evolutionary Algorithm (MOEA) was developed, taking into account that it is a well studied solution technique presenting good results for a large set of combinatorial optimization problems \cite{von2014survey}. The proposed MOEA is inspired in the Non-dominated Sorting Genetic Algorithm (NSGA-II) \cite{deb2012} and it mainly works in the following way (see Algorithm \ref{alg2:ga}):

The algorithm iterates over each set of requested VMs, that can be dynamically adjusted (or not) at each instant $t$. At step 1, a set $P_{0}$ of candidates is randomly generated. These candidates are repaired at step 2 to ensure that $P_{0}$ contains only feasible solutions. With the obtained non-dominated solutions, the first set $P_{known}$ (Pareto set approximation) is generated at step 3, considering lower and upper bounds associated to each objective function $z \in \{1, \dots, q\}$ $(L_z \leq f_z(x) \leq U_z)$. After initialization at step 4, evolution begins (between steps 5 and 12).

\begin{algorithm2e}[!t]
 \SetAlgoLined
 \KwData{CPU(t), MEM(t), PRC(t), AT(t), RT(t), n(t), H(t), P(t), selection strategy. See Section \ref{sec:problem_inputs} for notation details.}
 \KwResult{P(t+1). See Section \ref{sec:problem_outputs} for notation details.}
		initialize set of solutions $P_{0}$																\\
		$P_{0}'$ = repair infeasible solutions of $P_{0}$									\\
		update set of solutions $P_{known}$ from $P_{0}'$ applying lower and upper bounds	\\
		$u = 0 ; P_{u} = P_{0}'$																					\\
		\While{is not stopping criterion}{
			$Q_{u}$ = selection of solutions from $P_{u} \cup P_{known}$		\\
			$Q_{u}'$ = crossover and mutation of solutions of $Q_{u}$				\\
			$Q_{u}''$ = repair infeasible solutions of $Q_{u}'$							\\
			update set of solutions $P_{known}$ from $Q_{u}''$ applying lower and upper bounds \\
			increment number of generations $u$															\\
			$P_{u}$ = non-dominated sorting from $P_{u} \cup Q_{u}''$				\\
		}
		$P_{selected}$ = selected solution (selection strategy parameter)	\\
		\Return $P_{selected}$																						\\
		increment instant $t$; reset $P_{known}$													\\
 \caption{Proposed MOEA for multi-objective broker-oriented VMP.}
 \label{alg2:ga}
\end{algorithm2e}

The evolutionary process basically follows the same behavior: solutions are selected from the union of $P_{known}$ with the evolutionary set of solutions (or population) also known as $P_u$ (step 6), crossover and mutation operators are applied as usual (step 7), and eventually solutions are repaired, as there may be infeasible solutions (step 8). At step 9, the Pareto set approximation $P_{known}$ is updated (if applicable); while at step 10 the generation (or iteration) counter is updated. At step 11 a new evolutionary population $P_u$ is selected. The evolutionary process is repeated until the algorithm meets a stopping criterion (such as a maximum number of generations), returning one solution $P_{selected}$ from the set of solutions $P_{known}$ (step 13), using one of the strategies to be presented in Section \ref{sec2:selection_strategies}.

\subsection{Chromosome representation}
\label{sec2:chromosome_representation}

Considering the output data presented in Section \ref{sec:problem_outputs}, the proposed MOEA represents a current instance types selection and placement of VMs on available CSPs P(t) at instant $t$ as a chromosome. A chromosome (or solution representation of the proposed broker-oriented VMP problem) is represented as an integer matrix C(t) $\in \mathbb{Z}^{2 \times n(t)}$. The first row defines the selected instance types $(IT_j)$ for each requested VM, while the second row indicates the selected CSP $(c_k)$ in which the VM will be deployed. Note that the chromosome column size is variable, since $n(t)$ can change over time according to CST requirements.

\vspace{4mm}

\textbf{Example:} Suppose that a CST initially requires to deploy 13 VMs ($n(t) = 13$) at $t = 1$. There are three available CSPs ($m = 3$) with four different instance types ($l(t) = 4$) (see Table \ref{tab2:chroma}). Then, due to growing demands, the CST requires to increase number of deployed VMs ($n(t) = 16$) at $t = 2$ (see Table \ref{tab2:chromb}). For simplicity no reconfigurations of instance types or CSPs are presented in this very simple example.

\vspace{4mm}

According to Table \ref{tab2:chroma}, the obtained infrastructure configuration at $t = 1$ is:

\begin{itemize}
  \item $v_{1}$ is instance type $IT_2$ and located at CSP $c_1$;
  \item $v_{2}$ is type $IT_1$ and located at CSP $c_3$;
  \item $v_{3}$ is type $IT_3$ and located at CSP $c_1$;
	\item $v_{4}$ is type $IT_3$ and located at CSP $c_2$;
	\item $v_{5}$ is type $IT_1$ and located at CSP $c_2$;
  \item $v_{6}$ is type $IT_2$ and located at CSP $c_3$;
  \item $v_{7}$ is type $IT_4$ and located at CSP $c_1$;
	\item $v_{8}$ is type $IT_1$ and located at CSP $c_3$;
	\item $v_{9}$ is type $IT_4$ and located at CSP $c_3$;
	\item $v_{10}$ is type $IT_2$ and located at CSP $c_2$;
	\item $v_{11}$ is type $IT_2$ and located at CSP $c_1$;
	\item $v_{12}$ is type $IT_3$ and located at CSP $c_1$;
	\item $v_{13}$ is type $IT_3$ and located at CSP $c_2$.
\end{itemize}

When demand increases, three additional VMs are deployed ($v_{14}$ to $v_{16}$) totalizing 16 requested VMs at $t = 2$ ($n(t) = 16$), resulting in the infrastructure configuration presented in Table \ref{tab2:chromb}, where:

\begin{itemize}
	\item $v_{14}$ is type $IT_2$ and located at CSP $c_3$;
	\item $v_{15}$ is type $IT_3$ and located at CSP $c_2$;
	\item $v_{16}$ is type $IT_3$ and located at CSP $c_1$.
\end{itemize}

\begin{table}[t]
\centering
\begin{tabular}{|c|c|c|c|c|c|c|c|c|c|c|c|c|}
		\hline
		$v_1$ & $v_2$ & $v_3$ & $v_4$ & $v_5$ & $v_6$ & $v_7$ & $v_8$ & $v_9$ & $v_{10}$ & $v_{11}$ & $v_{12}$ & $v_{13}$ \\
    \hline
    2 & 1 & 3 & 3 & 1 & 2 & 4 & 1 & 4 & 2 & 2 & 3 & 3	\\
		\hline
		1 & 3 & 1 & 2 & 2 & 3 & 1 & 3 & 3 & 2 & 1 & 1 & 2	\\
		\hline
\end{tabular}
\caption{Example chromosome at $t = 1$.}
\label{tab2:chroma}
\end{table}

\begin{table}[t]
\centering
\begin{tabular}{|c|c|c|c|c|c|c|c|c|c|c|c|c|c|c|c|}
		\hline
		$v_1$ & $v_2$ & $v_3$ & $v_4$ & $v_5$ & $v_6$ & $v_7$ & $v_8$ & $v_9$ & $v_{10}$ & $v_{11}$ & $v_{12}$ & $v_{13}$ & $v_{14}$ & $v_{15}$ & $v_{16}$\\
    \hline
    2 & 1 & 3 & 3 & 1 & 2 & 4 & 1 & 4 & 2 & 2 & 3 & 3 & 2 & 3 & 3	\\
		\hline
		1 & 3 & 1 & 2 & 2 & 3 & 1 & 3 & 3 & 2 & 1 & 1 & 2 & 3 & 2 & 1	\\
		\hline
\end{tabular}
\caption{Example chromosome at $t = 2$.}
\label{tab2:chromb}
\end{table}

\begin{algorithm2e}[!b]
 \SetAlgoLined
 \KwData{Infeasible solution $C$}
 \KwResult{Feasible solution $C$}
		\While{$LOC_{min}$ constraint unfulfilled}{
			$c_{max}$ = select most loaded CSP from $C$													\\
			$c_{min}$ = select less loaded CSP from $C$													\\
			$v_{aux}$ = randomly select a VM currently located at $c_{max}$			\\
			$C$ = change $v_{aux}$ location from CSP $c_{max}$ to CSP $c_{min}$	\\
		}
		\Return Feasible solution $C$
 \caption{Infeasible Solution Reparation Algorithm.}
 \label{alg2:repairLOC}
\end{algorithm2e}

\subsection{Infeasible Solutions Reparation}
\label{sec2:infeasible_solutions_reparation}

With a random generation at the initialization phase (step 1 of Algorithm \ref{alg2:ga}) and/or solutions generated by standard genetic operators (step 7 of Algorithm \ref{alg2:ga}), infeasible solutions may appear, i.e. placement of VMs into CSPs which do not fulfill load balan\-cing constraint (see Section \ref{sec:load_balancing}). Repairing these infeasible solutions (steps 2 and 8 of Algorithm \ref{alg2:ga}) may be done as follows (see Algorithm \ref{alg2:repairLOC}): 

For each infeasible solution, the algorithm identifies the CSPs with the maximum and minimum number of VMs considering a particular CST request (step 2 and 3 respectively). Next, VMs are randomly changed from the most loaded CSP to the least loaded CSP, towards the accomplishment of the load balancing constraint. Finally, when the new infrastructure configuration meets $LOC_{min}$ constraint, the loop ends (step 6) and the repaired solution is returned (step 7).

\subsection{Variation Operators}
\label{sec2:variation_operators}

The proposed MOEA considers a \textit{Binary Tournament} method for selecting individuals for crossover and mutation operations \cite{coello2007}. The crossover operator implemented in the presented work is the single point cross-cut \cite{coello2007}. The selected individuals in the ascending population are replaced by individual in the descendant population.

This work considers a mutation method in which each gene is mutated with a probability $\frac{1}{n(t)}$, where $n(t)$ represents the number of requested VMs. This method offers the possibility of full uniform gene mutation, with a very low probability (but larger than zero), which is beneficial for search space exploration, reducing the probability of stagnation in a local optimum. The fitness function considered in the proposed algorithm is based on the one presented by Deb. et al. in \cite{deb2002fast}. 

The population evolution in the proposed MOEA is based on the population evolution proposed in the Non-dominated Sorting Genetic Algorithm II \cite{deb2002fast}. A new population $P_{u+1}$ is formed from the union of the best known population $P_{known}$ and offspring population $Q_{u}$, applying non-domination rank and crowding distance operators, as defined in \cite{deb2002fast}.

\subsection{Solutions Selection Strategies}
\label{sec2:selection_strategies}

Several challenges need to be addressed for dynamic formulations of broker-oriented VMP problems. In Pareto based algorithms, the Pareto set approximation can include a large number of non-dominated solutions; therefore, in a dynamic environment, automatically selecting only one of the non-dominated solutions (step 13 of Algorithm \ref{alg2:ga}) can be considered as a new difficulty for the problem. 

This work evaluates the following six selection strategies: (S1) random, (S2) minimum distance to origin, (S3) preferred solution, (S4) Maximum TICPU, (S5) Maximum TIMEM and (S6) Minimum TIP as is explained below.

\subsubsection{Random (\textit{S1})} Considering that the Pareto set approximation is composed by non-dominated solutions, randomly selecting one of the solutions could be an acceptable selection strategy.

\subsubsection{Minimum Distance to Origin (\textit{S2})} The solution with the minimum Euclidean distance to the origin is selected, considering all objective functions in a minimization context. For this purpose, $f_1(x)$ and $f_2(x)$ were redefined as the difference between the maximum possible value at instant $t$. When several solutions have equal Euclidean distance, only one of these solution is randomly selected.

\subsubsection{Preferred Solution (\textit{S3})} A solution is defined as preferred to another when it is better in more objective functions \cite{talavera2005policies}. When several solutions can be considered as preferred ones (there is a tie), the solution with higher TICPU or TIMEM function value is selected.

\vspace{2mm}

In order to be able to evaluate the proposed pure multi-objective formulation against mono-objective alternatives, this work considers selection strategies that selects the best solution according to only one objective, as the three described next.

\subsubsection{Maximum TICPU (\textit{S4})} This strategy select the solution with maximum TICPU function value. When several solutions have the same maximum TICPU value, only one of these solutions is randomly selected.

\subsubsection{Maximum TIMEM (\textit{S5})} This strategy select the solution with maximum TIMEM function value. When several solutions have the same maximum TIMEM value, only one of these solutions is randomly selected.

\subsubsection{Minimum TIP (\textit{S6})} This strategy select the solution with minimum TIP function value. When several solutions have the same minimum TIP value, only one of these solutions is randomly selected.

\section{Experimental Results}
\label{sec:experimental_results}

The main goal of the proposed experiments is to evaluate selection strategies to automatically select one solution from a Pareto set approximation at each time instant as well as evaluating the scalability of the proposed algorithm when solving problem instances with large number of requested VMs. This section summarizes obtained results and main findings of this work.

\subsection{Experimental Environment}
\label{sec:experimental_environment}

The proposed MOEA (see Section \ref{sec:MOEA}) was implemented using Java programming language. Experimental scenarios include real input data from different CSPs, consisting in several instances types in terms of CPU [in \# of cores] and RAM [in GB] (see Table \ref{tab2:instancetypes}), as well as prices [in USD] (see Table \ref{tab2:prices}) and allocation and release times [in seconds] (see Table \ref{tab2:allocation}). The source code of the implemented algorithm as well as input data and experimental results are available online\footnote{https:$\slash\slash$github.com$\slash$lgchamorro$\slash$broker-VMP-MOEA}.

\begin{table}[htbp]
\centering
\begin{tabular}{|c|c|c|c|c|c|}
		\hline
		\textbf{Resources} & \textbf{m} & \textbf{s} & \textbf{M} & \textbf{L} & \textbf{XL}	\\
    \hline
    $CPU (Cores)$ & 1 & 1 & 2 & 2 & 4 \\
		\hline
		$Memory (GB)$ & 1 & 2 & 4 & 8 & 16 \\
		\hline
\end{tabular}
\caption{VMs hardware configuration per instance type. Micro(m), Small(s), Medium(M), Large(L), Extra Large(XL).}
\label{tab2:instancetypes}
\end{table}

\begin{table}[htbp]
\centering
\begin{tabular}{|c|c|c|c|c|c|}
		\hline
		\textbf{CSP} & \textbf{m} & \textbf{s} & \textbf{M} & \textbf{L} & \textbf{XL}	\\
    \hline
    $EC2-US$ & 0.013 & 0.026 & 0.052 & 0.104 & 0.239 \\
		\hline
		$EC2-EU$ & 0.014 & 0.028 & 0.056 & 0.112 & 0.264 \\
		\hline
		$EC2-OC$ & 0.020 & 0.040 & 0.080 & 0.160 & 0.336 \\
		\hline
\end{tabular}
\caption{Hourly instance type prices per CSP [in \$]. Micro(m), Small(s), Medium(M), Large(L), Extra Large(XL).}
\label{tab2:prices}
\end{table}

\begin{table}[htbp]
\centering
\begin{tabular}{|c|c|c|c|c|c|c|c|c|c|c|}
		\hline
		  & \multicolumn{5}{c}{\textit{Allocation}} & \multicolumn{5}{|c|}{\textit{Release}} \\
		\hline
		\textbf{CSP} & \textbf{m} & \textbf{s} & \textbf{M} & \textbf{L} & \textbf{XL} & \textbf{m} & \textbf{s} & \textbf{M} & \textbf{L} & \textbf{XL} \\
    \hline
    $EC2-US$ & 71 & 82 & 85 & 90 & 64 & 20 & 21 & 20 & 20 & 25 \\
		\hline
		$EC2-EU$ & 71 & 82 & 85 & 90 & 64 & 20 & 21 & 20 & 20 & 25 \\
		\hline
		$EC2-OC$ & 71 & 82 & 85 & 90 & 64 & 20 & 21 & 20 & 20 & 25 \\
		\hline
\end{tabular}
\caption{Allocation and release times per instance type [in seconds]. Micro(m), Small(s), Medium(M), Large(L), Extra Large(XL).}
\label{tab2:allocation}
\end{table}

Table \ref{tab2:othersinput} summarizes parameters related to the proposed MOEA. The parameters are: $t$ represents the time period in which CSPs maintain th same offers and CSTs maintain the same requirements (for practical reasons, the value of $t$ is constant); the proposed MOEA was was executed several times (\# of runs) in order to average all the obtained results and finally the population parameters (population quantity and generations).

\begin{table}[htbp]
\centering
\begin{tabular}{|c|c|}
		\hline
		\textbf{Parameter} & \textbf{Value}	\\
		\hline
		$t$ & 24 hours											\\
    \hline
		\# of runs & 10										\\
		\hline
		Population Size & 50								\\
		\hline
		Number of Generations & 200					\\
		\hline
\end{tabular}
\caption{MOEA general parameters considered in experiments.}
\label{tab2:othersinput}
\end{table}

CSTs can define minimum and maximum (depending of the optimization context) values that they expect for each considered objective function (also knows as minimum or maximum acceptable required values) as presented in Tables \ref{tab2:reqscenario01} and \ref{tab2:reqscenario02}. Also, CSTs may define a tolerance margin (in percent over specific expected value) within which an output value could be considered as valid. Both, expected values and tolerance percentage are optional problem inputs defined by CSTs. Output values are evaluated against expected values in a optimization context. This means that according to the objective function, the expected value could designate an acceptable minimum (TICPU, TIMEM) or maximum (TIP). In the same way, tolerance could extend below minimum acceptable values (TICPU, TIMEM) or above the maximum acceptable value (TIP) based on the tolerance percentage against expected value indicated by CSTs.

\begin{table}[htbp]
\centering
\begin{tabular}{|c|c|c|c|c|c|c|}
		\hline
		 \multicolumn{2}{|c}{ }  & \multicolumn{3}{|c|}{\textbf{Expected}} & \textbf{Restriction} & \textbf{Tolerance} \\
		\hline
		\textbf{$t$} & \textbf{$VMs$} & \textit{TICPU} & \textit{TIMEM} & \textit{TIP} & $LOC_{min}$ in \% & Margin in \% \\
    \hline
		1 & 100 & 300 & 1300 & 26 & 30 & 10	\\
		\hline
		2 & 100 & 300 & 1300 & 26 & 30 & 10	\\
		\hline
		3 & 120 & 380 & 1400 & 30 & 30 & 10	\\
		\hline
		4 & 120 & 380 & 1400 & 30 & 30 & 10	\\
		\hline
		5 & 120 & 380 & 1400 & 30 & 30 & 10	\\
		\hline
		6 & 120 & 380 & 1400 & 30 & 30 & 10	\\
		\hline
		7 & 100 & 300 & 1300 & 26 & 30 & 10	\\
		\hline
\end{tabular}
\caption{CST requirements for Experiment 1.}
\label{tab2:reqscenario01}
\end{table}

\begin{table}[htbp]
\centering
\begin{tabular}{|c|c|c|c|c|c|c|}
		\hline
		 \multicolumn{2}{|c}{ } & \multicolumn{3}{|c|}{\textbf{Expected}} & \textbf{Restriction} & \textbf{Tolerance} \\
		\hline
		\textbf{$t$} & \textbf{$VMs$} & \textit{TICPU} & \textit{TIMEM} & \textit{TIP} & $LOC_{min}$ in \% & Margin in \% \\
    \hline
		1 & 400 & 1100 & 4300 & 100 & 30 & 10	\\
		\hline
		2 & 400 & 1100 & 4300 & 100 & 30 & 10	\\
		\hline
		3 & 500 & 1400 & 5100 & 130 & 30 & 10	\\
		\hline
		4 & 500 & 1400 & 5100 & 130 & 30 & 10	\\
		\hline
		5 & 500 & 1400 & 5100 & 130 & 30 & 10	\\
		\hline
		6 & 500 & 1400 & 5100 & 130 & 30 & 10	\\
		\hline
		7 & 400 & 1100 & 4300 & 100 & 30 & 10	\\
		\hline
\end{tabular}
\caption{CST requirements for Experiment 2.}
\label{tab2:reqscenario02}
\end{table}

Experiments for each of the six evaluated selection strategies were repeated 10 times, given the probabilistic nature of Evolutionary Algorithms (EAs). Two experiments were performed considering different scenarios. For simplicity, each instant $t$ represents 24 hours, as described below:

\vspace{2mm}

\textbf{Experiment 1:} 

\begin{itemize}
  \item $t = 1$ (Day 1). Initial infrastructure placement. Requested VMs $n(t) = 100$.
	\item $t = 2$ (Day 2). Changes in CSP offers. New instance type offer (micro instances).
	\item $t = 3$ (Day 3). Changes in CST requirements. CST needs more VMs $n(t) = 120$.
	\item $t = 4$ (Day 4). Changes in CSP offers. Nighttime instance type discounts. The $EC2-OC$ provider offers 50\% off.
	\item $t = 5$ (Day 5). Changes in CSP offers. Nighttime instance type discounts ends.
	\item $t = 6$ (Day 6). Changes in CSP offers. An instance type is removed from the market (xlarge instances).
	\item $t = 7$ (Day 7). Changes in CST requirements. CST needs less VMs $n(t) = 100$.
\end{itemize}

\textbf{Experiment 2:} 

\begin{itemize}
  \item $t = 1$ (Day 1). Initial infrastructure placement. Requested VMs $n(t) = 400$.
	\item $t = 2$ (Day 2). Changes in CSP offers. New instance type offer (micro instances).
	\item $t = 3$ (Day 3). Changes in CST requirements. CST needs more VMs $n(t) = 500$.
	\item $t = 4$ (Day 4). Changes in CSP offers. Nighttime instance type discounts. $EC2-OC$ provider offers 50\% nighttime off.
	\item $t = 5$ (Day 5). Changes in CSP offers. Nighttime instance type discounts ends.
	\item $t = 6$ (Day 6). Changes in CSP offers. An instance type is removed from the market (xlarge instances).
	\item $t = 7$ (Day 7). Changes in CST requirements. CST needs less VMs $n(t) = 400$.
\end{itemize}

\begin{table}[!bp]
\centering
\begin{tabular}{|c|c|c|c|c|}
		\hline
		 \multicolumn{2}{|c}{ } & \multicolumn{3}{|c|}{\textit{Average}} \\
		\hline
		\textbf{No.} &\textbf{Selection Strategy} & \textbf{$TICPU$} & \textbf{$TIMEM$} & \textbf{$TIP$} \\
    \hline
		S1 & Random					& 2,502.69 & 9,474.74 & 165.09 		\\
		\hline
		S2 & Distance				& 2,543.16 & 9,696.35 & 169.75 		\\
		\hline
		S3 & Preferred			& 2,725.89 & 10,556.18 & 184.84 	\\
		\hline
		S4 & Maximum TICPU	& 2,711.18 & 10,476.18 & 183.67 	\\
		\hline
		S5 & Maximum TIMEM	& 2,712.62 & 10,501.05 & 183.87 	\\
		\hline
		S6 & Minimum TIP		& 2,315.17 & 8,567.42 & 148.20 		\\
		\hline
\end{tabular}
\caption{Obtained averaged results for Experiment 1.}
\label{tab2:outputscenario01}
\end{table}

\begin{table}[!bp]
\centering
\begin{tabular}{|c|c|c|c|c|}
		\hline
		 \multicolumn{2}{|c}{ } & \multicolumn{3}{|c|}{\textit{Average}} \\
		\hline
		\textbf{No.} &\textbf{Selection Strategy} & \textbf{$TICPU$} & \textbf{$TIMEM$} & \textbf{$TIP$} \\
    \hline
		S1 & Random & 8,767.46 & 31,615.79 & 547.48 					\\
		\hline
		S2 & Distance & 8,926.14 & 32,362.23 & 561.90 				\\
		\hline
		S3 & Preferred & 9,137.18 & 33,369.80 & 579.78 				\\
		\hline
		S4 & Maximum TICPU & 9,096.16 & 33,132.56 & 576.34 		\\
		\hline
		S5 & Maximum TIMEM & 9,107.74 & 33,233.36 & 576.87 		\\
		\hline
		S6 & Minimum TIP & 8,433.01 & 29,995.62 & 518.41 			\\
		\hline
\end{tabular}
\caption{Obtained averaged results for Experiment 2.}
\label{tab2:outputscenario02}
\end{table}

\subsection{Selection Strategy Analysis}
\label{sec:selection_strategy_results}

Since the focus of this work is the simultaneous optimization of all three objective functions in a PMO context, a comparison is made considering the concept of Pareto dominance. Considering that six selection strategies were evaluated, these strategies are compared through two methods \cite{ihara2015} in order to analyse which selection strategy could be considered as the best optionin terms of the values of the objective functions: 

\begin{itemize}
	\item Pareto Dominance: a solution $u_1$ dominates another $u_2$ if considering each objective function $u_1$, is better or equal than $u_2$ and strictly better in at least one objective.
	\item Pareto Preference: a solution $u_1$ is defined as preferred over another $u_2$ if $u_1$ has more objective functions better evaluated than $u_2$ in terms of quantity.
\end{itemize}

Tables \ref{tab2:outputscenario01} and \ref{tab2:outputscenario02} summarize obtained results in both experiments. As can be seen in Tables \ref{tab2:scenario01dominance} and \ref{tab2:scenario02dominance} (Pareto dominance column), in average none of the strategies is dominated by another strategy in both experiments; consequently, no strategies dominates the others, i.e. we can not establish that a given strategy is better than the other.

\begin{table}[ht]
\centering
\begin{tabular}{|c|c|c|c|c|c|c|c|}
		\hline
		 \multicolumn{2}{|c}{ } & \multicolumn{6}{|c|}{\textit{Pareto Dominance}} \\
		\hline
		\textbf{No.} &\textbf{Selection Strategy} & $S1$ & $S2$ & $S3$ & $S4$ & $S5$ & $S6$ \\
    \hline
		S1 & Random 				& N/A & - & - & - & - & - 			\\
		\hline
		S2 & Distance 			& - & N/A & - & - & - & - 			\\
		\hline
		S3 & Preferred 			& - & - & N/A & - & - & - 			\\
		\hline
		S4 & Maximum TICPU	& - & - & - & N/A & - & - 			\\
		\hline
		S5 & Maximum TIMEM	& - & - & - & - & N/A & -				\\
		\hline
		S6 & Minimum TIP		& - & - & - & - & - & N/A 			\\
		\hline
\end{tabular}
\caption{Experiment 1. Pareto dominance relation for the evaluated strategies.}
\label{tab2:scenario01dominance}
\end{table}

\begin{table}[ht]
\centering
\begin{tabular}{|c|c|c|c|c|c|c|c|}
		\hline
			\multicolumn{2}{|c}{ } & \multicolumn{6}{|c|}{\textit{Pareto Preference}}\\
		\hline
		\textbf{No.} &\textbf{Selection Strategy} & $S1$ & $S2$ & $S3$ & $S4$ & $S5$ & $S6$ \\
    \hline
		S1 & Random 				& N/A & - & - & - & - & $\succ$ 													\\
		\hline
		S2 & Distance 			& $\succ$ & N/A & - & - & - & $\succ$											\\
		\hline
		S3 & Preferred 			& $\succ$ & $\succ$ & N/A & $\succ$ & $\succ$ & $\succ$		\\
		\hline
		S4 & Maximum TICPU	& $\succ$ & $\succ$ & - & N/A & - & $\succ$								\\
		\hline
		S5 & Maximum TIMEM 	& $\succ$ & $\succ$ & - & $\succ$ & N/A & $\succ$					\\
		\hline
		S6 & Minimum TIP		& - & - & - & - & - & N/A																	\\
		\hline
\end{tabular}
\caption{Experiment 1. Pareto preference relation for the evaluated strategies.}
\label{tab2:scenario01preference}
\end{table}

Given that none of the considered strategies can be declared as the best strategy considering exclusively Pareto Dominance, a further comparison of selection strategies using the preference method (i.e. more quantity of better objective functions values) criteria \cite{von2014survey} is presented in Tables \ref{tab2:scenario01preference} and \ref{tab2:scenario02preference}. 

It may seem intuitive that the $S3$ strategy (that uses the preference criterion itself) should be the best; as experimentally validated and presented in Tables \ref{tab2:scenario01preference} and \ref{tab2:scenario02preference} in both experiments, given that on each evaluated sceanrio, the preferred selection strategy (S3) selects the solution containing the greatest number of best objective functions.

\begin{table}[htp]
\centering
\begin{tabular}{|c|c|c|c|c|c|c|c|}
		\hline
		  \multicolumn{2}{|c}{ } & \multicolumn{6}{|c|}{\textit{Pareto Dominance}} \\
		\hline
		\textbf{No.} & \textbf{Selection Strategy} & $S1$ & $S2$ & $S3$ & $S4$ & $S5$ & $S6$ \\
    \hline
		S1 & Random					& N/A & - & - & - & - & -	\\
		\hline
		S2 & Distance				& - & N/A & - & - & - & -	\\
		\hline
		S3 & Preferred			& - & - & N/A & - & - & -	\\
		\hline
		S4 & Maximum TICPU	& - & - & - & N/A & - & -	\\
		\hline
		S5 & Maximum TIMEM	& - & - & - & - & N/A & -	\\
		\hline
		S6 & Minimum TIP		& - & - & - & - & - & N/A	\\
		\hline
\end{tabular}
\caption{Experiment 2. Pareto dominance relation for the evaluated strategies.}
\label{tab2:scenario02dominance}
\end{table}

\begin{table}[htp]
\centering
\begin{tabular}{|c|c|c|c|c|c|c|c|}
		\hline
		  \multicolumn{2}{|c}{ } & \multicolumn{6}{|c|}{\textit{Pareto Preference}}\\
		\hline
		\textbf{No.} &\textbf{Selection Strategy} & $S1$ & $S2$ & $S3$ & $S4$ & $S5$ & $S6$ \\
    \hline
		S1 & Random					& N/A & - & - & - & - & $\succ$								\\
		\hline
		S2 & Distance				& $\succ$ & N/A & - & - & - & $\succ$					\\
		\hline
		S3 & Preferred			& $\succ$ & $\succ$ & N/A & $\succ$ & $\succ$ & $\succ$ \\
		\hline
		S4 & Maximum TICPU	& $\succ$ & $\succ$ & - & N/A & - & $\succ$		\\
		\hline
		S5 & Maximum TIMEM	& $\succ$ & $\succ$ & - & $\succ$ & N/A & $\succ$ \\
		\hline
		S6 & Minimum TIP		& - & - & - & - & - & N/A												\\
		\hline
\end{tabular}
\caption{Experiment 2. Pareto preference relation for the evaluated strategies.}
\label{tab2:scenario02preference}
\end{table}

\section{Conclusion and Future Works}
\label{sec:conclusions}

Current cloud computing markets have dynamic environment where the providers offers variability about pricing schemes and computational resources, and the CSTs requirement may change over time (e.g. available budget, VM resources demands). In this context, the broker-oriented VMP problem resolution represents a true challenge.

State of the art research has studied this problems in a mono-objective approach. Chamorro et al. \cite{chamorro2016} proposes a genetic algorithm for optimizing a single objective function focused on scalability issues raised in previous works.  

This work proposes for first time a formulation for broker-oriented VMP problem resolution in a pure multi-objective context. This formulation simultaneously optimize three objective functions: (i) TICPU, (ii) TIMEM and (iii) TIP, subject to load balancing across CSPs. Also it includes a fully dynamic environment where VM resources offers and pricing from CSPs are changing, and CSTs have variables requirements in terms of resources and budget.

In order to solve the proposed multi-objective broker-oriented VMP problem, a Multi-Objective Evolutionary Algorithm (MOEA) was developed that is able of effectively solve large-scale instances to the proposed formulation of the problem.

In Pareto based algorithms, the Pareto set can include several non-dominated solutions but only one can be used for the reconfiguration of the new CST infrastructure. This work evaluates the following six solution selection strategies: (i) random, (ii) minimum distance to origin, (iii) preferred solution, (iv) Maximum TICPU, (v) Maximum TIMEM and (vi) Minimum TIP. 

Several experiments were performed to assess the MOEA performance against large instances of the problem and to evaluate the quality of the obtained solutions. The experiments were focused on two scenarios which comprises a dynamic environment through resources offers and pricing by CSPs and requirements of CST, as well as large instances of the problem (up to 500 VMs). They proved that the algorithm obtains good solutions that meet CSTs requirements. Between the selection strategies, preferred solution strategy has showed that gets the best solution in terms of CSTs requirements.

As future works, the authors suggest to study the heterogeneity that may exist between different providers, generally in terms of computational resources. In addition, researchers could perform a study on the feasibility of implementing a predictive model of the behavior of CSPs offers and CSTs requirements in a purely multi-objective context.



\newcommand{\newblock}{}

\bibliographystyle{unsrt}

\bibliography{main}

\begin{thebibliography}{10}

\bibitem{buyya2008market}
Rajkumar Buyya, Chee~Shin Yeo, and Srikumar Venugopal.
\newblock Market-oriented cloud computing: Vision, hype, and reality for
  delivering it services as computing utilities.
\newblock In {\em High Performance Computing and Communications, 2008. HPCC'08.
  10th IEEE International Conference on}, pages 5--13. Ieee, 2008.

\bibitem{LucasSimarro2011}
Jos{\'e}~Luis Lucas~Simarro, Rafael Moreno-Vozmediano, Ruben~S Montero, and
  Ignacio~Mart{\'\i}n Llorente.
\newblock Dynamic placement of virtual machines for cost optimization in
  multi-cloud environments.
\newblock In {\em High Performance Computing and Simulation (HPCS), 2011
  International Conference on}, pages 1--7. IEEE, 2011.

\bibitem{lopez2015}
F.~L\'opez-Pires and B.~Bar\'an.
\newblock A virtual machine placement taxonomy.
\newblock In {\em Cluster, Cloud and Grid Computing (CCGrid), 2015 15th
  IEEE/ACM International Symposium on}, pages 159--168, May 2015.

\bibitem{Tordsson2012}
Johan Tordsson, Rub{\'e}n~S Montero, Rafael Moreno-Vozmediano, and Ignacio~M
  Llorente.
\newblock Cloud brokering mechanisms for optimized placement of virtual
  machines across multiple providers.
\newblock {\em Future Generation Computer Systems}, 28(2):358--367, 2012.

\bibitem{von2014survey}
Christian von L{\"u}cken, Benjam{\'\i}n Bar{\'a}n, and Carlos Brizuela.
\newblock A survey on multi-objective evolutionary algorithms for
  many-objective problems.
\newblock {\em Computational Optimization and Applications}, 58(3):707--756,
  2014.

\bibitem{chaisiri2009}
Sivadon Chaisiri, Bu-Sung Lee, and Dusit Niyato.
\newblock Optimal virtual machine placement across multiple cloud providers.
\newblock In {\em Services Computing Conference, 2009. APSCC 2009. IEEE
  Asia-Pacific}, pages 103--110, 2009.

\bibitem{Mark2011}
Ching Chuen~Teck Mark, Dusit Niyato, and Tham Chen-Khong.
\newblock Evolutionary optimal virtual machine placement and demand forecaster
  for cloud computing.
\newblock In {\em Advanced Information Networking and Applications (AINA), 2011
  IEEE International Conference on}, pages 348--355. IEEE, 2011.

\bibitem{Kessaci2013}
Yacine Kessaci, Nouredine Melab, and E-G Talbi.
\newblock A pareto-based genetic algorithm for optimized assignment of vm
  requests on a cloud brokering environment.
\newblock In {\em Evolutionary Computation (CEC), 2013 IEEE Congress on}, pages
  2496--2503. IEEE, 2013.

\bibitem{Amato2013}
Alba Amato, Beniamino Di~Martino, and Salvatore Venticinque.
\newblock Cloud brokering as a service.
\newblock In {\em P2P, Parallel, Grid, Cloud and Internet Computing (3PGCIC),
  2013 Eighth International Conference on}, pages 9--16. IEEE, 2013.

\bibitem{Amato2013a}
Alba Amato and Salvatore Venticinque.
\newblock Multi-objective decision support for brokering of cloud sla.
\newblock In {\em Advanced Information Networking and Applications Workshops
  (WAINA), 2013 27th International Conference on}, pages 1241--1246. IEEE,
  2013.

\bibitem{Li2011}
Wubin Li, Johan Tordsson, and Erik Elmroth.
\newblock Modeling for dynamic cloud scheduling via migration of virtual
  machines.
\newblock In {\em Cloud Computing Technology and Science (CloudCom), 2011 IEEE
  Third International Conference on}, pages 163--171. IEEE, 2011.

\bibitem{chamorro2016}
Lino Chamorro, Fabio L\'opez-Pires, and Benjam\'in Bar\'an.
\newblock A genetic algorithm for dynamic cloud application brokerage.
\newblock {\em IEEE International Conference on Cloud Engineering}, 2016.

\bibitem{amazon2015}
Amazon~Web Services.
\newblock Amazon ec2 pricing schemes.
\newblock {\em https://aws.amazon.com/es/ec2/pricing}, Jul 2012.

\bibitem{coello2007}
Carlos~Coello Coello, Gary~B Lamont, and David A~Van Veldhuizen.
\newblock Evolutionary algorithms for solving multi-objective problems.
\newblock {\em Springer}, 2007.

\bibitem{amazon2009}
Amazon~Web Services.
\newblock Announcing micro instances for amazon ec2.
\newblock {\em
  https://aws.amazon.com/es/about-aws/whats-new/2010/09/09/announcing-micro-instances-for-amazon-ec2},
  Sep 2009.

\bibitem{amazon2014}
Amazon~Web Services.
\newblock New low cost ec2 instances with burstable performance.
\newblock {\em
  https://aws.amazon.com/es/blogs/aws/low-cost-burstable-ec2-instances/}, Jul
  2014.

\bibitem{amazon2016}
Amazon~Web Services.
\newblock Cloud services pricing.
\newblock {\em https://aws.amazon.com/es/pricing/services/}, Jul 2016.

\bibitem{iosup2011performance}
Alexandru Iosup, Simon Ostermann, M~Nezih Yigitbasi, Radu Prodan, Thomas
  Fahringer, and Dick Epema.
\newblock Performance analysis of cloud computing services for many-tasks
  scientific computing.
\newblock {\em IEEE Transactions on Parallel and Distributed systems},
  22(6):931--945, 2011.

\bibitem{flopezmany2015}
L\'opez-Pires Fabio and Benjam\'in Bar\'an.
\newblock A many-objective optimization framework for virtualized datacenters.
\newblock {\em 5th International Conference on Cloud Computing and Service
  Science}, 2015.

\bibitem{deb2012}
Sameer~Agarwal Kalyanmoy~Deb, Amrit~Pratap and T.~Meyarivan.
\newblock A fast and elitist multiobjective genetic algorithm: Nsga-ii.
\newblock {\em IEEE Transactions on Evolutionary Computation}, 2012.

\bibitem{deb2002fast}
Kalyanmoy Deb, Amrit Pratap, Sameer Agarwal, and TAMT Meyarivan.
\newblock A fast and elitist multiobjective genetic algorithm: {NSGA-II}.
\newblock {\em Evolutionary Computation, IEEE Transactions on}, 6(2):182--197,
  2002.

\bibitem{talavera2005policies}
Francisco Talavera, Jorge Crichigno, and Benjam{\'\i}n Bar{\'a}n.
\newblock Policies for dynamical multiobjective environment of multicast
  traffic engineering.
\newblock In {\em IEEE ICT}, 2005.

\bibitem{ihara2015}
Diego Ihara, Fabio L\'opez-Pires, and Benjam\'in Bar\'an.
\newblock Many-objective virtual machine placement for dynamic environments.
\newblock {\em International Conference on Utility and Cloud Computing}, 2015.

\end{thebibliography}

\end{document}